\documentstyle[11pt,twoside,graphicx]{article}

\newcommand{\kms}{\hbox{\,km\,s$^{-1}$}}
\def\arcsec{\hbox{$^{\prime\prime}$}}
\def\sun{\hbox{$\odot$}}

\begin{document}

\footnotesize{
{\it \noindent Eta Carinae: Reading the Legend}\\
{\it Mt. Rainier; 10.-13.July 2002}\\
{\it Editor: Bruce Balick}} \\

\begin{center}
\vspace*{1cm}
\Large{\bf The Physical Structure of the Outer Ejecta
and the Strings}\\
\vspace*{0.5cm}
\large{\bf Kerstin Weis\footnotemark} \\
\footnotetext{Feodor-Lynen-fellow, Alexander-von-Humboldt foundation}
\medskip
\parbox{14cm}{
\normalsize{Max-Planck-Institut f\"ur Radioastronomie, Auf dem H\"ugel 69, 53121
  Bonn, Germany and University of Minnesota, 116 Church Street SE,
  Minneapolis, MN 55455, USA }}

\vspace*{1cm}

\parbox{14cm}{
\section*{Abstract}
\normalsize
The outer ejecta is part of the nebula around $\eta$ Carinae. 
They are filamentary, shaped irregularly and larger than the Homunculus, 
the central bipolar
nebula. While the Homuculus is mainly a reflection nebula, the outer ejecta
is an emission structure. 
However, we showed with kinematic analysis that 
the outer ejecta (as the Homunculus) 
expands bi-directional despite of its complex morphology. 
Radial velocities in the 
outer ejecta reach up to 2000\,\kms\ and give 
rise to X-ray emission. An analysis showing the distribution of the soft X-ray
emission and its comparison to the optical emitting gas is presented here.
X-ray maxima are found in areas in which the expansion velocities are
highest. The temperature of 0.65 keV determined with the CHANDRA/ACIS data 
and thermal equilibrium models indicates post-shock velocities of 750\,\kms ,
about what was found in the spectra. In addition analysis of the new
HST-STIS data from the Strings---long, highly collimated structures in the
outer ejecta---are presented. The data show that the electron 
density of the Strings is of the order of 10$^4$\,cm$^{-3}$. The same value
was detected for other structures in the outer ejecta. 
With this density String 1 has a mass of about 3 10$^{-4}$\,M$_{\sun}$ and the
total ejecta could be as massive as 0.5 M$_{\sun}$.}

\end{center}
\vspace*{1cm}

\textwidth17cm
\normalsize
\section{The outer ejecta: Introduction and background}

\begin{figure}
\includegraphics[width=8.4cm]{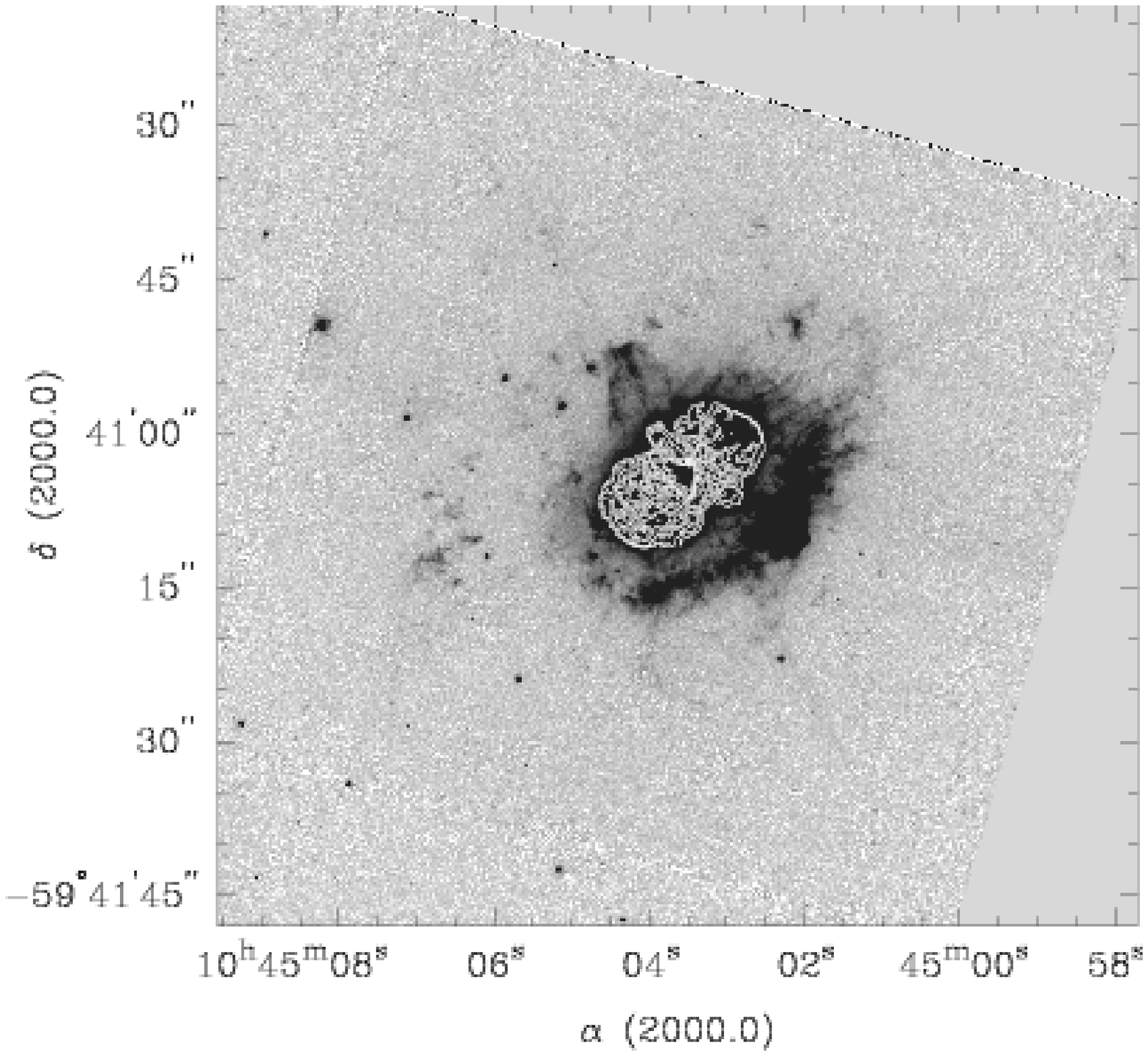}\vspace{0.3cm}
\includegraphics[width=8.4cm]{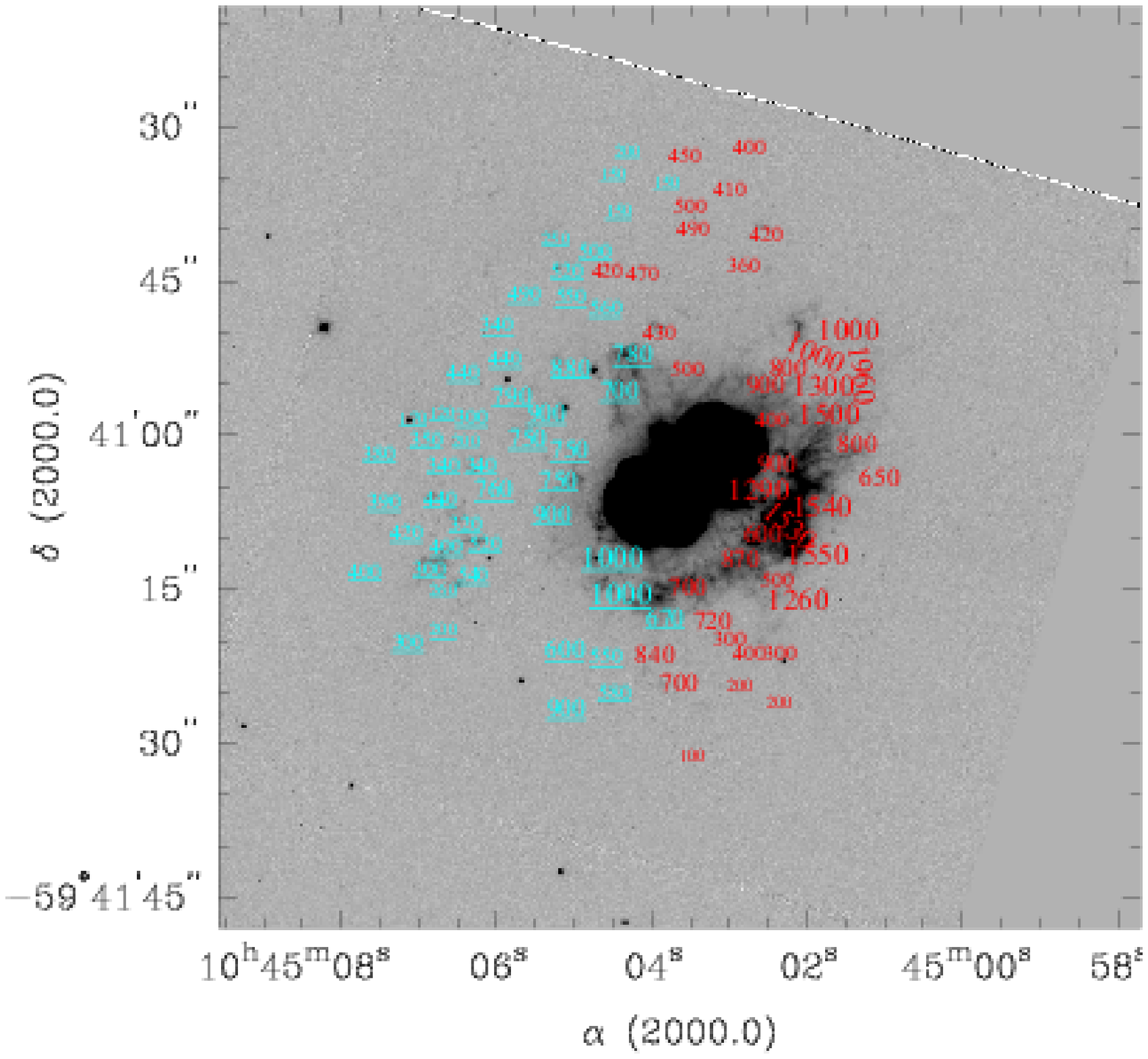}
\caption{{\it Left:} An HST/F658N image of the nebula around $\eta$ Carinae, 
the brightest regions, the Homunculus, are additionally overplotted 
with intensity contours,
for a better illustration and comparison. {\it Right:} Same HST image but 
with radial velocities overplotted at certain regions. The font size is
larger for larger velocities. Red and blue (plus underlined) colors indicate
red- (positive) and blueshifted (negative) radial velocities.   
}\label{figure1}
\end{figure}

The first images of the nebula around $\eta$ Carinae were made by 
Gaviola (1946, 1950) and Thackeray (1949, 1950). Gaviola named the nebula
according to the geometry he identified at that time the {\it Homunculus}.
It had a size of somewhat larger than 10\arcsec.
Nowadays we know that the Homunculus is highly symmetric---bipolar---and 
is larger. The Homunculus is only the central part of the nebula 
around $\eta$ Carinae (e.g.  Walborn 1976), which in total extends 
to a diameter of 60\arcsec\ (0.67\,pc). While we still call the central 
bipolar structure the
Homunculus, all outer filaments are combined into what is known as the 
{\it  outer ejecta}. The sizes and morphology of structures in the outer
ejecta are manifold. In contrary to the Homunculus is the outer ejecta not
symmetric, nor is it a coherent object but consists out of quite a number of 
filaments, bullets or knots. 
Fig. \ref{figure1} (left) shows an HST image taken with the F658N filter 
of the nebula around $\eta$ Carinae. Besides the different morphology of the 
Homunculus and the outer ejecta, it illustrates also the large difference in
brightness. Therefore the Homunculus is additionally plotted with contours.
If instead we compare the kinematics of Homunculus and outer ejecta, they seem
more alike. The Homuculus, according to his bipolarity expands with about
650\,\kms (Davidson \& Humphreys 1997, Currie et al. 1996) and with the 
south-east lobe approaching and the north-west 
lobe receeding. The outer ejecta has on average similar velocities,
the majority of the structures expanding with $|v_{\rm exp}|$ = 600\,\kms
(e.g. Meaburn et al. 1996, Weis et al. 2001a).
But note that a significant fraction of the filaments move much faster,
they reach velocities as high as 2000\,\kms (Weis 2001a,b). 
So in velocity space nevertheless the outer ejecta are more ordered 
than expected from the
morphology. In the south-eastern
region most filaments are blueshifted, while in the north-west the clumps are 
redhifted, see right panel in Fig. \ref{figure1}. Compared to the movement of the Homunculus the expansion  
of the outer ejecta is along a very similar symmetry axis. As the Homunculus
the outer ejecta has a bi-directional (bipolar) movement. 

\section{X-ray emission from the outer ejecta}

\begin{figure}
\includegraphics[width=8.1cm]{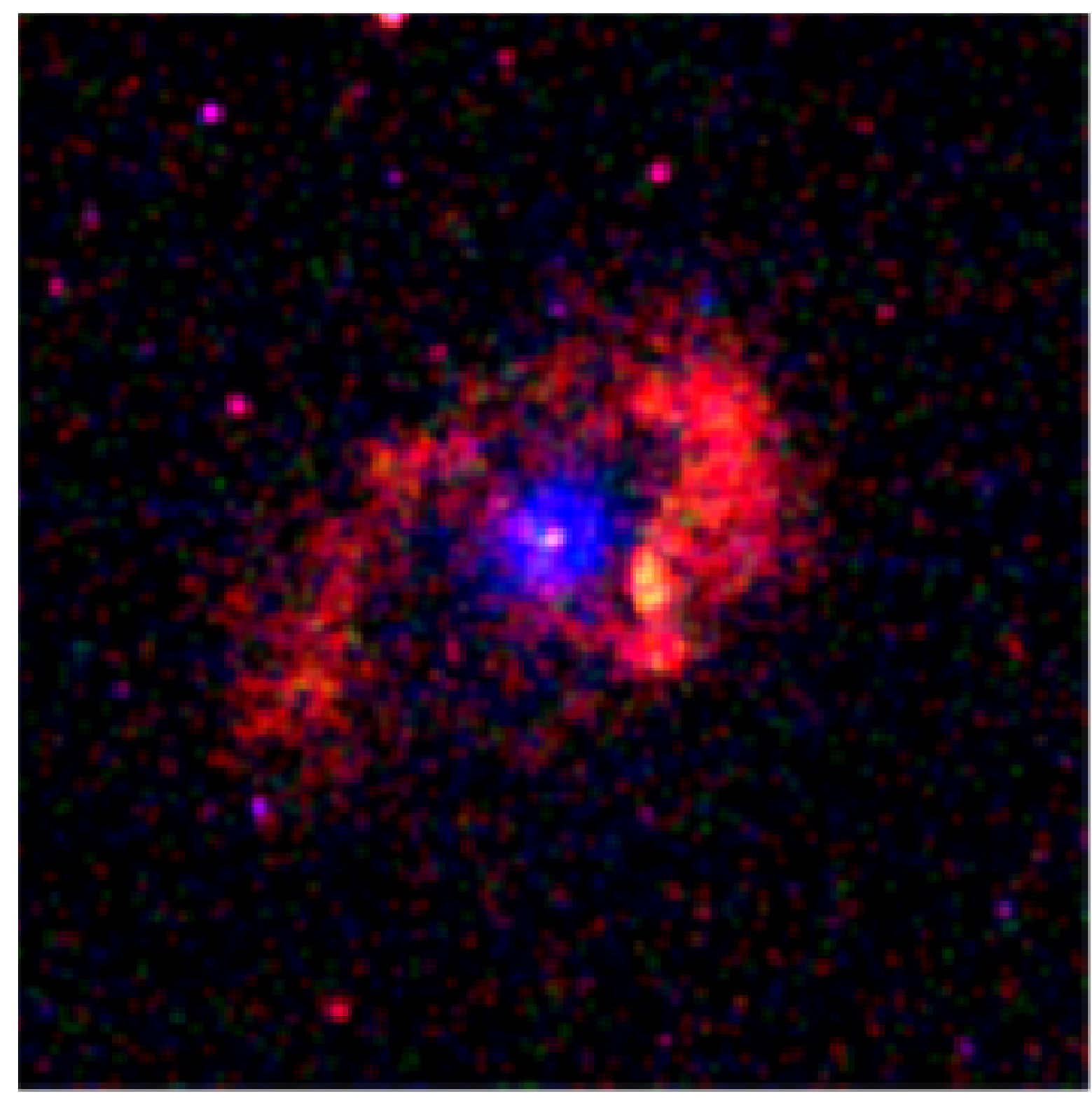}\vspace{0.3cm}
\includegraphics[width=8.7cm]{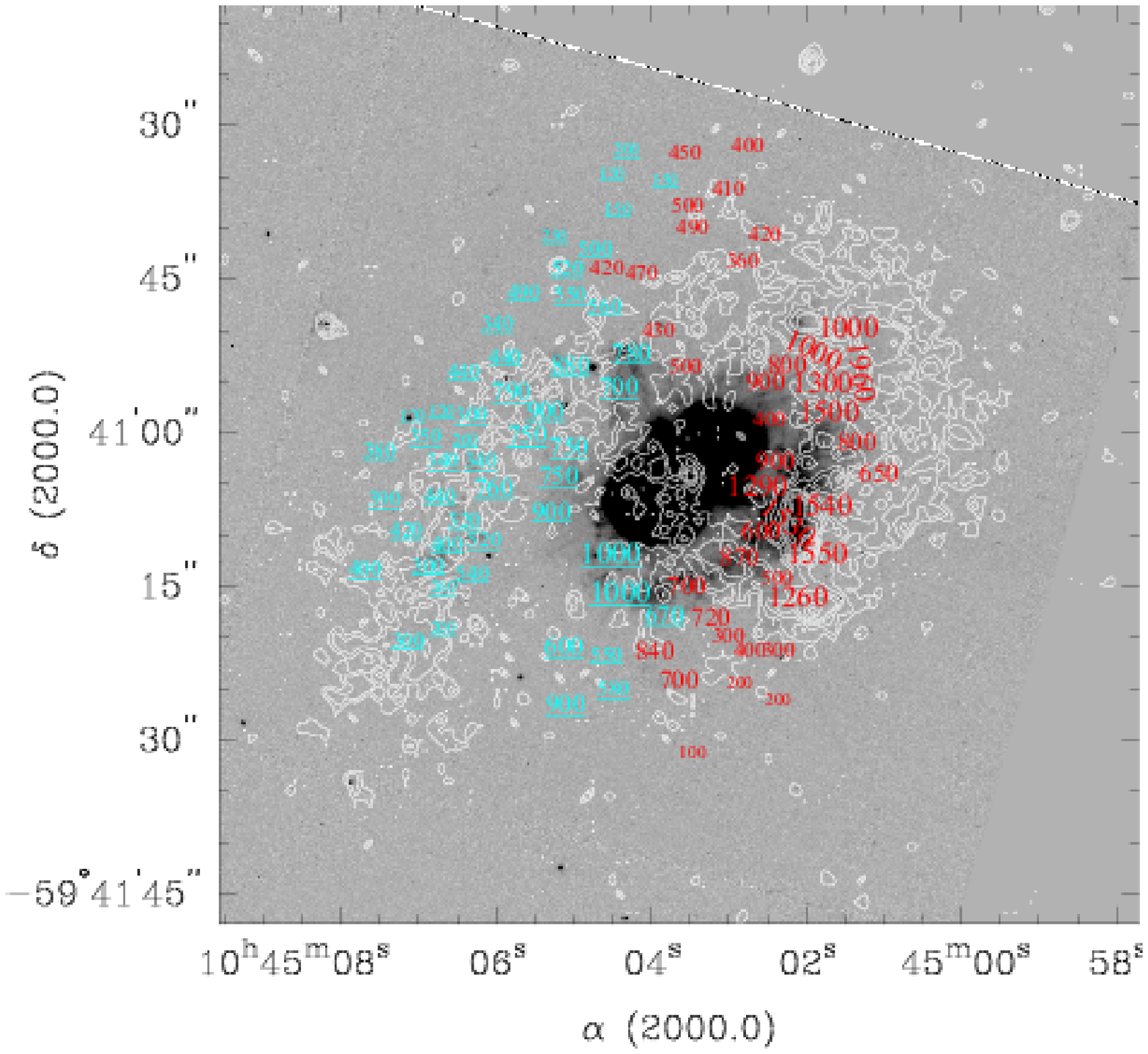}\caption{{\it Left:} This figure
  shows a color composite image of $\eta$ Carinae and its nebula in
X-rays. The images was taken with CHANDRA/ACIS. Color coding: Green
0.2-0.6\,keV (predominantly the Nitrogen line at 0.5\,keV); 
Red 0.6-1.2\,keV; Blue 1.2-11\,keV. {\it Right:} The CHANDRA 
emission in the 0.6-1.2\,keV regime overlayed on the HST/F658N image. 
The softer emission agrees with the outer ejecta. Again the expansion
velocities are added, and the highest expansion velocities are situated 
on X-ray emission maxima. 
}\label{figure2}
\end{figure}

As expected from the expansion velocities, the nebula around $\eta$ Carinae is
deteced in X-rays, as is the central object. Originally barely resolved with
the EINSTEIN satellite (Chlebowski et al.\ 1984) the
higher spatial resolution of ROSAT and CHANDRA makes it possible to better
separate the emission of the nebula from that of the harder central source.
Already with ROSAT it was found that the softer emission is roughly 
hook shaped (e.g. Corcoran et al. 1994, Weis
et al.\ 2001) and agrees in its dimensions with the extension of the 
outer ejecta (Weis et al.\ 2001).
The CHANDRA/ACIS data are so far the images with the highest spatial 
resolution and at the same time had decent spectral resolution. In Figure 
\ref{figure2} (left) this image is shown color-coded with red is the soft
emission and blue the hard emission. 
The central source---which might be slightly extended---is very hard and 
as known from other observations variable (e.g. Corcoran et al.\ 1995, 
Ishibashi et al.\ 1999, Weis et al.\ 2001). The emission 
which results from the outer ejecta is much softer and
shows a hook shape, plus a bridge like connection which crosses the central
object from the south-west end of the hook to the ridge in the 
north-east. This is best seen in the contours shown in Fig. \ref{figure2} 
(right). Still with CHANDRAs higher resolution 
an overlay of the optical image and the X-ray emission (Fig. \ref {figure2};
right) shows only a few  correlations between the 
optically emitting and the hot X-ray gas.
While the dimensions of the outer ejecta in the optical and X-ray 
are roughly the same, individual knots rarely agree. One exception is the so
called {\it S\,condensation} (Walborn 1976), here the brightest X-ray feature agrees
with the bright optical filament.   
A much stronger conformance was found comparing the image with the radial 
expansion velocities. The 
expansion velocities are derived from our optical Echelle spectra
(see Weis 2001a,b) and are plotted in Fig. \ref{figure2} (right).
We find that areas with higher X-ray emission, show in general 
higher expansion velocities, so the faster the gas is moving 
the more intense is the X-ray emission found in that region. 
Therefore we can explain the morphology of the extended soft X-ray emission
from $\eta$ Carinae by shocks formed through the interaction of the 
faster moving filaments. 
Since CHANDRA/ACIS-S has good energy resolution we have the possibility to 
extracte spectra. A spectrum of the emission in
connection with the outer ejecta was extracted and modelled.
Using a one temperature, thermal equilibrium model (Mewe-Kaastra-Liehdal)
and a fixed column density
(${\rm log\,N}_{\rm H} \sim$ 21.3), yielded already reasonable
good fits to the data, with a temperature of $\sim$ 0.65 keV 
(Weis et al. 2002, Weis et al. in prep.). That 
indicates post-shock velocities of 
around 750\,\kms\ and agrees quite well with the detected average expansion
velocities. Besides the temperature, the nitrogen abundance was determined,
which was possible because a very prominent nitrogen line at
about 0.5\,keV was visible in all spectra. In general the values indicate a
roughly 6 times higher nitrogen abundance. A similar enhancement of Nitrogen 
was detected in optical spectra of the outer ejecta. 

\section{HST-STIS observations of the Strings---first results}

\begin{figure}
\includegraphics[width=7.5cm]{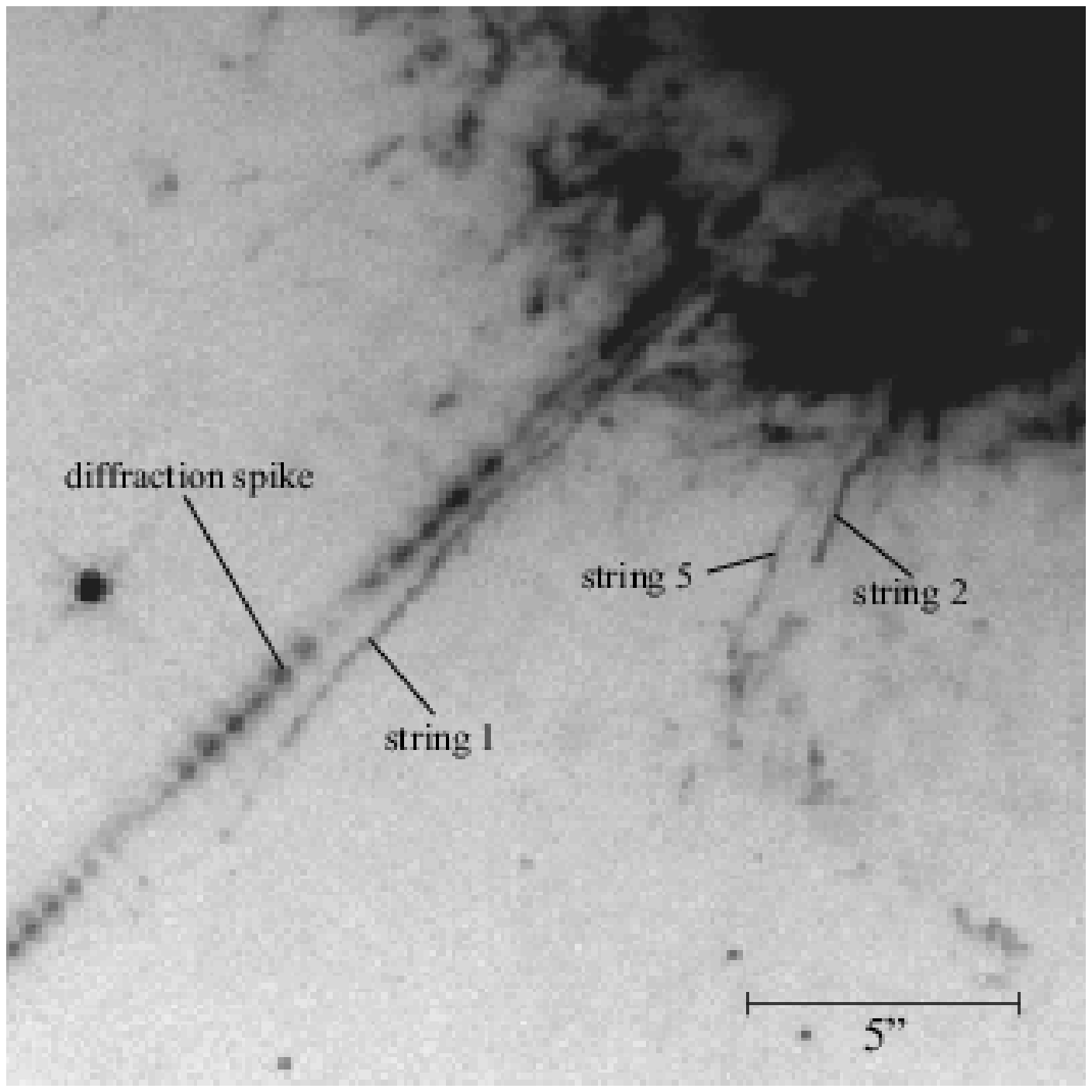}\vspace{0.3cm}
\includegraphics[width=9.2cm]{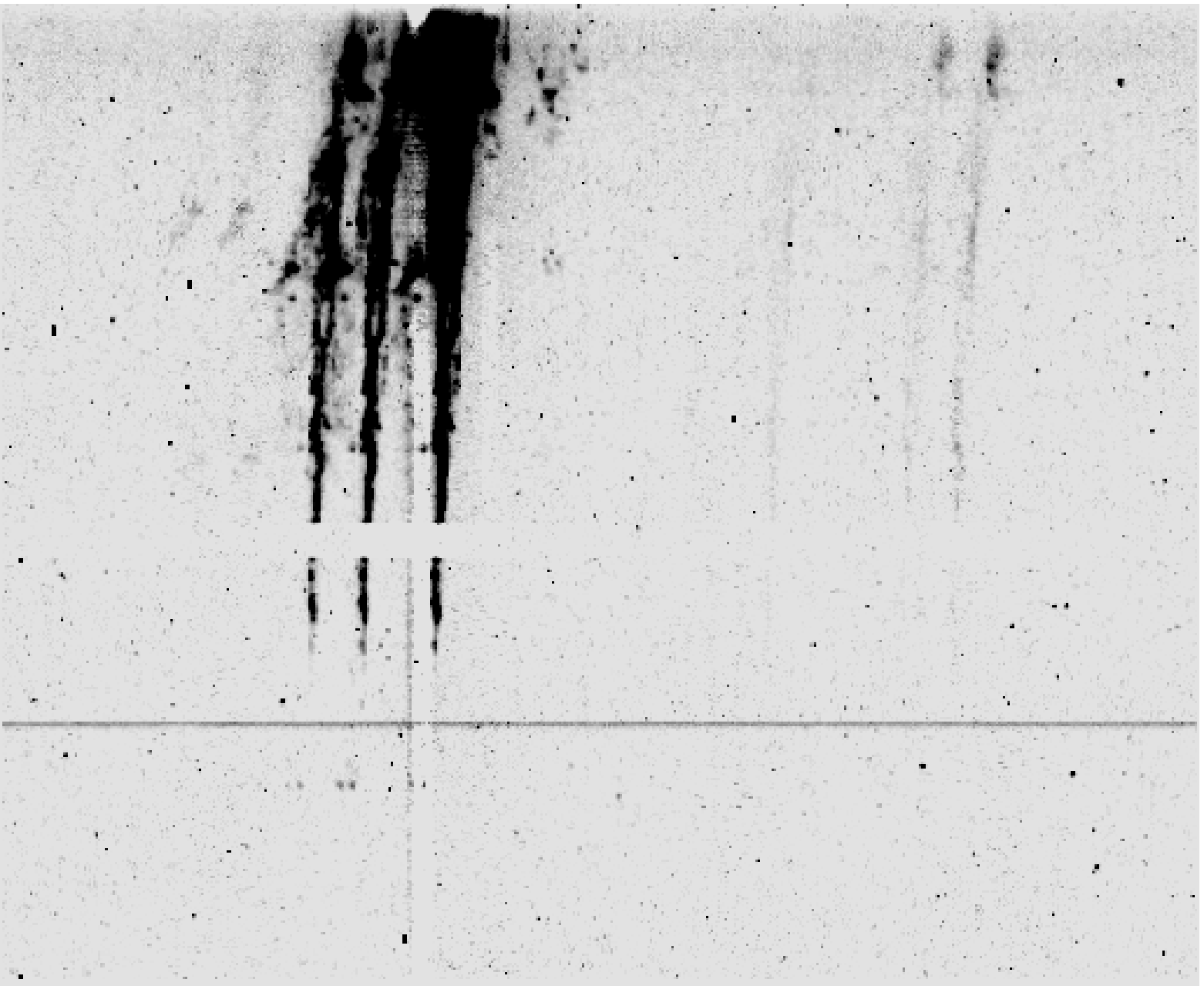}
\caption{{\it Left:} This image shows a section of the  HST/F656N image 
of $\eta$ Carinae in the region were three of the 5 identified Strings are. 
String\,1 is the longest of all. {\it Right:} Part of the HST-STIS spectrum of 
String\,1  between 6450 and 6780\,\AA.}\label{figure3}
\end{figure}

The Strings are long, highly collimated structures that point nearly 
radially away from $\eta$ Carinae and emerge from the outer ejecta 
region that is closest to the Homunculus, partly that is the {\it S\,ridge}
(Walborn 1976).
The largest of all is String\,1 (Fig.\ \ref{figure3} with 
a total length 0.177\,pc and a width of equal or less than 0.003\,pc). 
String\,4 (0.13\,pc), String\,3 (0.085\,pc), String\,5 (0.058\,pc), and 
String\,2 (0.044\,pc) follow, all with similar
width. For a detailed description of the basic parameters of the Strings see
Weis et al.\ (1999). The Strings are nearly aligned with the long axis of the
Homunculus and their radial velocity increases as the String move outwards.
For String\,1 this increase ranges from $-522$ to about $-995$\,\kms (Weis et al.\
1999). Here I will discuss new results from a first analysis of the HST-STIS
spectra taken of 4 of the Strings; due to the space limitation only the case
of String\,1 is discussed in detail.\\

\noindent We observed the Strings with the HST-STIS and the G750M first 
order grating and the 52\,$\times$\,0.5 aperture. The slit was successfully
aligned with each
String. The spectra ranged from 6295 to 6867\,\AA. In all spectra the Strings
are identified in emission in the following lines: [N\,{\sc ii}] at 6548\,\AA\
and 6583\,\AA, H$_{\alpha}$, He\,{\sc i} at 6678\,\AA\ and both [S\,{\sc ii}]
 lines at 6716 and 6731\,\AA. Figure \ref{figure3} shows String\,1 in these
 lines. This imaging-spectroscopic display of String\,1 shows that the String
 is very homogeneous, is not subdivided and that there is no brighter structure
 at the front. The bullets originally believed to be the head of String\,1 are
 visible and appear detached from the String---below the continuum of the
 star in Fig. \ref{figure3} (right)---nevertheless they follow 
the linear velocity increase well. 
From these spectra it is obvious that the String is more of a steady flow
rather than bullets along a chain. 
The origin of String\,1 is sudden, within the southern part of the S ridge. 
It looks as it does not extend back into the Homunculus, see top of
Fig.\,\ref{figure3} right panel.
The velocities determined from the STIS spectra range from $-290$\,\kms\ at the
beginning to $-950$\,\kms\ further out. The slightly different values 
compared to Weis et al.\ (1999) result
mainly because of the lower spectral resolution of STIS ($\sim$ 50\,\kms\ 
compared to 14\,\kms) and as for the much
lower starting value due to the better tracing of the string inwards giving
the higher spatial resolution. We therefore trace the String longer than in
our Echelle spectra.
Also a split which was suggested to occur in String\,1 (Weis et al.\ 1999) 
was confirmed with the STIS data, see Figs.\,\ref{figure3}.
From the lines detected we derived line ratios for String\,1, see table 1. 
The line ratios are very similar for string 2-4, the other Strings observed.  
Using the [S\,{\sc ii}] line ratio and assuming a temperature of about 
14\,000\,K (a typical value for the S ridge, Dufour et al.\ 1997) 
we obtain an electron density of about 1.2 
10$^4$ cm$^{-3}$. Since the ratio lies close to the high density limit this 
is more of a conservative lower limit and might be higher. The results
agree well with electron densities in the outer ejecta (S\,ridge)
determined by Dufour et al. (1997) using the Si\,{\sc iii}] lines.  
With the help of the electron density and kinematics of String\,1 one can
calculate the mass and determine the kinetic energy associated with the
String. With an electron density of about 10$^4$\,cm$^{-3}$, which 
corresponds to a mass density $\rho$ of 1.6 10$^{-23}$\,kg\,cm$^{-3}$ 
(using cosmic abundances) and 
a volume of the String of 3.7 10$^{49}$\,cm$^{3}$, the mass $M$ 
of String\,1 is 
about 3 10$^{-4}$\,M$_{\sun}$ if it is completely filled. 
With a lower limit on the average expansion velocity 
of $v\sim$450\,\kms\ the total kinetic 
energy $E=1/2\,M\,v^2$ is about 6 10$^{43}$\,ergs. 
Or using the cross section $A$ 
of the String and its density flow we can estimate the power in sense of a
kinetic Luminosity $L=1/2\,\rho\,v^2\,A\,v$ which is about 1\,L$_{\sun}$.
Several mechanism have been proposed to explain the Strings, like trails left
by a bullet (Weis et al.\ 1999, Redman et al.\ 2002), shadowing effects 
(Soker 2001) or a steady gas flow (Weis et al.\ 1999) are only a few to name. 
The new STIS data show that the Strings are dense (but might be hollow)
objects with 10$^4$\,cm$^{-3}$ and that there is no density contrast detected
along the String which could give rise to a denser leading head. 
There still might be a density difference
to much higher densities which we would not detect since we are close to the
high density limit of the [S\,{\sc ii}] line ratio. But this is unprobable
since than at the same time the surface brightness should also
increase in this denser knot, and this is not see in the spectra.

\begin{table}[t]
\caption[]{Line ratios of String\,1}\label{table1}
\begin{center}
\begin{tabular}{ccccc}
\hline
[S\,{\sc ii}] & [N\,{\sc ii}]/H$_{\alpha}$ & [S\,{\sc
  ii}]/H$_{\alpha}$ & He\,{\sc i}/H$_{\alpha}$ & He\,{\sc i}/[S\,{\sc ii}]\\
I(6716)/I(6731) & I(6583)/I(6563) & I(6716+31)/I(6563) & I(6678)/I(6563)
& I(6678)/I(6716+31)\\
\hline
0.5 & 1.7-3 & 0.07 & 0.03 & 0.5 \\
\hline
\end{tabular}
\end{center}
\end{table}

\section{Summary and general remarks}

The nebula around $\eta$ Carinae can be divided into two quite different
parts, the Homunculus a bipolar reflection nebula about 0.2\,pc across and the
0.67\,pc large outer ejecta, a filamentary emission nebula. We have shown that
the kinematics of the outer ejecta indicate that also this part of the nebula
expands bi-directional, or bipolar. The symmetry axis is similar to that of the
Homunculus. \\Historically we know that $\eta$ Carinae und its nebula emit 
X-rays. 
With CHANDRAs unprecedented spatial resolution a very accurate comparison of
the X-ray emission with the optical emission could be made. The X-ray emission 
is more homogeneous and smoother than the optical nebula, which is more of a 
accumulation of filaments, knots and bullets. While the sizes match, that is 
we find X-ray emission in regions were there is optical emission we barely
find an agreement of individual knots, except for the S condensation. 
Comparing the intensity maxima of the X-ray emission   with the velocities of
the optical filaments yields a much better consensus. We conclude that 
in the case of $\eta$ Carinae's outer ejecta the faster moving 
filaments are able to form stronger shocks and therefore stronger X-ray 
emission. The temperature of 0.65\,keV indicates post-shock velocities
of 750\,\kms, in agreement with the measurements.  \\ 
From the new HST-STIS spectra of the Strings we obtain a first analysis of new
parameters of the Strings. We can see that the Strings starts abruptly and
does not extend back into the Homunculus. The slowest velocity of String\,1
was denoted with $-290$\,\kms. The fastest is similar  to the previous 
measurements with about $-950$\,\kms. We determined several line ratios
for String\,1, the most interesting of which is the [S\,{\sc ii}] ratio, a
density indicator. For String\,1 this ratio is about 0.5 $\pm$ 0.1. Within the
errors the ratio is steady along the String. The ratio is close---but clearly
not always at---the high
density limit of the line, so we still can determine an electron density 
of 10$^4$\,cm$^{-3}$. The density, as the line ratios in general are 
for all Strings alike. The Strings are more of a denser steady flow rather
than ablating bullets, shadows or knots on chains.
Since the HST-STIS spectra we have taken for the Strings also contain
information of the immediate surrounding---that is the outer ejecta, mainly
the S ridge and {\it W\,Arc}---we could also determine the density of several
filaments in this region. Measurements in these spectra show that the 
filaments in the outer ejecta have a density of the order of
10$^4$\,cm$^{-3}$, thus about the same as the Strings. This
value does not change significantly---at least in the regions which are
covered by out HST-STIS observations. If we assume that all filaments in the
outer ejecta have roughly this density, and take a reasonable filling 
factor for the knots in the ejecta of 1\% within the 
total (0.67)$^3$\,pc$^3$ volume 
of the outer ejecta, the total mass is at least 0.5\,M$_{\sun}$.\\

{\bf  \noindent Acknowledgments:} The author thanks Michael F. Corcoran and 
Kris
Davidson for help and discussion on the CHANDRA data and 
Ted Gull for providing a much-better-than-pipeline HST-STIS data reduction 
and advice on technical aspects concerning these data. Special thanks go to 
Dominik Bomans for discussion on
the subject and commending on the manuscript.\\

\section*{References}
\begin{verse}

Chlebowski, T., Seward, F. D., Swank, J., Szymkowiak, A., 1984, 
ApJ, 281, 665 \\
Corcoran, M. F., Swank J., Rawley, G., Petre, R., Schmitt J., Day, C. 1994,
AIP Conf. Series, 313, 159 \\ 
Corcoran, M. F., Rawley, G. L., Swank, J. H., Petre, R. 1995, ApJ, 445, L121\\ 
Currie, D., Dowling, D. M., Shaya, E. J., et al. 1996, AJ, 112, 1115\\
Davidson, K., Humphreys, R. M., 1997, ARA\&A, 35, 1\\
Dufour, R.,  Glover, T. W., Hester, J. J., Currie, D. G., van Orsow, D., 
Walter, D. K. 1997, in Luminous Blue Variables: Massive Stars in Transition,
ASP Conf. Series; Vol. 120, ed. Antonella Nota \& Henny Lamers, 255 \\    
Gaviola, E. 1946, Revista Astronomica, 18, 25\\
Gaviola, E. 1950, ApJ, 111, 408\\
Ishibashi, K., Corcoran, M. F., Davidson, K., et al.\ 1999, ApJ, 524, 983\\ 
Meaburn, J., Boumis, P., Walsh, J. R., et al., 1996, MNRAS, 282, 1313\\
Redman, M. P., Meaburn, J., Holloway, A. J. 2002, MNRAS, 332, 754\\
Soker, N. 2001, A\&A, 377, 672\\
Thackeray, A. D. 1949, Observatory, 69, 31\\
Thackeray, A. D. 1950, MNRAS, 110, 524\\
Walborn, N. R. 1976, ApJ, 204, L17\\
Weis, K., 2001a, in $\eta$ Carinae and other mysterious stars,
ed. T. Gull, S. Johanson, \& K. Davidson, ASP Conf. 242, 129 \\
Weis, K., 2001b, in 
Reviews in Modern Astronomy 14, Springer-Verlag, 
ed.: Schielicke, R.E, ISBN 3-9805176-4-0, 261 \\
Weis, K., Duschl, W. J., Chu, Y.-H., 1999,A\&A, 349, 467  \\
Weis, K., Duschl, W. J., Bomans, D. J., 2001, A\&A, 367, 566 \\
Weis, K., Corcoran, M. F., Davidson, K., Humphreys, R.M. 2002, in: K.A. 
van der Hucht, A. Herrero \& C. Esteban (eds.), A Massive Star Odyssey, from 
Main Sequence to Supernova, Proc. IAU Symp. No. 212 (San Francisco: ASP), 
in press

\end{verse}
\end{document}